
\magnification 1200
\font\ftitle=cmbx10 scaled\magstep2

\centerline{\ftitle Stochastic bosonization in arbitrary dimensions}
\bigskip\bigskip\bigskip
\centerline {\bf L. Accardi\qquad Y.G.
Lu\footnote{$^{(*)}$}{\rm Dipartimento di Matematica, Universit\`a di Bari.}
\qquad I. Volovich\footnote{$^{(**)}$}{\rm On leave absence from Steklov
Mathematical Institute, Vavilov St. 42, GSP--1, 117966, Moscow.}}
\bigskip\bigskip\bigskip
\centerline {\bf Centro V. Volterra}\smallskip
\centerline{\bf Universit\`a di Roma ``Tor Vergata''}\smallskip
\centerline{\bf via di Tor Vergata,}\smallskip
\centerline{\bf 00133 -- Roma}\smallskip
\centerline{E--Mail volterra@mat.utovrm.it}
\bigskip\bigskip\bigskip

\noindent{\it Abstract\/}. A procedure of bosonization of Fermions in
an arbitrary dimension is suggested. It is shown that a quadratic
expression in the fermionic fields after rescaling time $t\to t/\lambda^2$
and performing the limit $\lambda\to0$ (stochastic limit), gives rise to
a bosonic operator satisfying the boson canonical commutation relations. This
stochastic bosonization of Fermions is considered first for free fields
and then for a model with three--linear couplings. The limiting dynamics
of the bosonic theory turns out to be described by means of a quantum
stochastic differential equations.\bigskip\bigskip


In a previous paper [AcLuVo] we have found the following
direct relation between quantum theory,
in real time, and stochastic processes:
$$\lim_{\lambda\to0}\lambda A\left({t\over\lambda^2}\,,\
k\right)=B(t,k)\eqno(1.1)$$
where $A(t,k)$ is a free dynamical evolution of a usual (Boson or Fermion)
annihilation
operator $a(k)$ in a Fock space ${\cal F}$, i.e.
$$A(t,k)=e^{itE(k)}a(k)\eqno(1.2)$$
$k$ is a momentum variable and $B(t,k)$ is a Boson or Fermion quantum
field acting in another Fock space ${\cal H}$ and
satisfying the canonical commutation relations:
$$[B(t,k),B(t',k')]_\pm=2\pi\delta(t-t')\delta(k-k')\delta(E(k))\eqno(1.2a)$$
(cf. [AcLuVo] for a precise explanation of this result). Notice that, in
(1.1), $\lambda $ can also be interpreted as the square root of Planck's
constant: in this case the limiting relation (1.1) can be interpreted as
a new type of semi--classical expansion, describing not the first order
approximation to the classical solution, but rather the {\it
fluctuations} around it.

Here $\lambda$ can be interpreted as the Planck of semiclassical
expansion. The relation () is similar to (1.9) and $\lambda$ plays the
role fo $1/\sqrt N$.

A (Boson or Fermion) Fock field satisfying commutation relations of the
form (1.2a) is called a {\it quantum white noise\/} [Ac Fri Lu1] and is a
prototype example of quantum stochastic process.
In terms of the field operators $\phi(t,\vec x)$ one can rewrite (1.1) as
$$\lim_{\lambda\to0}\lambda\phi\left({t\over\lambda^2}\,,\
x\right)=W(t,x)\eqno(1.3)$$
and one can prove that the vacuum correlations of
$W(t,x)$ coincide with those of a {\it classical Brownian motion\/}.

The scaling $t\mapsto t/\lambda^2$ has its origins in the early attempts
by Pauli, Friedrichs and Van Hove to deduce a {\it master equation\/}
and in the line of research originated from these papers it was used to deduce
the effects of the interaction of one or more atoms with a field (or gas)
on the dynamics of the atoms.

In a series of papers starting from [AcFrLu] (cf. the introduction of
for an historical survey)
it has been shown that it is also possible to control the limiting
dynamics of the quantum fields themselves under the scaling
$t\to t/\lambda^2$ and to prove that the quantum fields converge to quantum
Brownian motions and the Heisenberg equation to a quantum Langevin
equation. Moreover, as shown in a multiplicity of quantum systems, involving
the basic physical interactions, this is a rather universal phenomenon.

This kind of limit and the set of mathematical techniques developed to
establish it, was called in [AcLuVo] {\it the stochastic limit of quantum
field theory\/}. We call it also the $1+3$ asymptotical expansion to
distinguish it from analogous scalings generalizing the present one in
the natural direction of rescaling other variables beyond (or instead
of) time: space, energy particle density,...

The main result of this paper is a generalization of (1.1) to a pair of
{\it Fermi operators\/} $A,A^+$ and it can be
described by the formula
$$\lim_{\lambda\to0}\lambda A\left({t\over\lambda^2}\,,\ k_1\right)A
\left({t\over\lambda^2}\,,\ k_2\right)=B(t,k_1,k_2)\eqno(1.4)$$
The remarkable property of the formula (1.4) is that while the $A(t,k)$
are {\bf Fermion} annihilation operators the limit field $B(t,k_1,k_2)$ is a
{\bf Boson}
annihilation operator which satisfies with its hermitian
conjugate, the canonical bosonic commutation relations. For this
reasons we call the
formula (1.4): {\it stochastic bosonization\/}.

The bosonization of Fermions is well known in $1+1$ dimensions, in
particular in the Thirring--Luttinger model see, for example [Wh],
[SV] and in string theory. The stochastic bosonization (1.4)
takes place in the real $1+3$ dimensional space--time and in fact in any
dimension.
For previous discussion of bosonization in higher dimensions cf. [Lut],
[Hal].

In particular in quantum chromodynamics one can think of the
Fermi--operators $A(t,k)$ as corresponding to quarks then the bosonic
operator $B(t,k_1,k_2)$ can be considered as describing a meson (cf.
[AcLuVo qcd] for a preliminary approach to QCD in this spirit).

In the following section we proove the formula (1.4) and then in Section
3 we consider the stochastic bosonization of a model with nontrivial
interaction.\bigskip\medskip

\noindent{\bf \S 2.) Boson Fock spaces as stochastic limits of Fermion Fock
spaces}

\bigskip
Let $\Gamma_-({\cal H}_1)$ denote the Fermi Fock space on the
1--particle space ${\cal H}_1$ and let, for $f\in{\cal H}_1$, $A(f)$,
$A^+(f)$ denote the creation and annihilation operators on
$\Gamma_-({\cal H}_1)$ which satisfy the usual canonical anticommutation
relations (CAR):
$$A(f)A^+(g)+A^+(g)A(f)=\langle f,g\rangle\eqno(2.1)$$
The main idea of stochastic bosonization is that, in a limit to be specified
below, two Fermion operators
give rise to a Boson operator. To substantiate the idea
let us introduce the operators
$${\cal A}(f,g):=A(g) A(f),\qquad {\cal A}^+(f,g) := \bigl({\cal A}(f,g)
\bigr)^*\eqno(2.2)$$
then by the CAR we have that
$${\cal A}(f,g){\cal A}^+(f',g')=A(g) A(f)A^+(f') A^+(g')=$$
$$=<g,g'><f,f'>-<f,f'><g,g'>+{\cal A}^+(f',g'){\cal A}(f,g)+R(f,g;f',g')
\eqno(2.3)$$
i.e.
$$[{\cal A}(f,g),{\cal A}^+(f',g') ]= <(f\otimes g),\ (f'\otimes
g')>+R(f,g;f',g')
\eqno(2.4)$$
where we introduce the notations
$$R(f,g;g',f'):=<f,g'>A^+(f')A(g) -<f,f'>A^+(g')A(g)
-<g,g'>A^+(f')A(f)\eqno(2.5a)$$
$$<(f\otimes g),\ (f'\otimes g')> :=<f,f'><g,g'>-<f,g'><g,f'> \eqno(2.5b)$$
Moreover,
$${\cal A}(f,g){\cal A}(f',g')  ={\cal A}(f',g'){\cal A}(f,g)
\eqno(2.6)$$

In the following the identities (2.4) and (2.6) will be called the {\it
quasi--CCR\/}. We shall prove that, in the stochastic limit, the {\it
remainder term\/} $R$ tends to zero so that, in this limit, the {\it
quasi CCR\/} become {\it bona fide\/} CCR.

The first step in the stochastic limit of quantum field theory is to
introduce the {\it collective operators\/}. In our case we associate to
the quadratic Fermion operator (2.2)
the collective creation operator defined by:
$$A^+_\lambda(S,T;f_0,f_1):=\lambda\int_{S/\lambda^2} ^{T/\lambda^2}
e^{i\omega t} A^+(S_tf_0)A^+(S_tf_1)dt\eqno(2.7)$$
where $\omega$ is a real number (whose physical meaning explained at
the end of Section (3.),
$S\le T$, $f_0,f_1\in {\cal K}$ and $S_t:{\cal H}_1\to{\cal H}_1$
is the one--particle dynamical
evolution whose second quantization gives the free evolution of the
Fermi fields, i.e.
$$A(f)\mapsto A(S_tf)\ ;\quad A^+(f)\mapsto A^+(S_tf)$$

The collective annihilation
operators are defined as the conjugate of the collective creators.

The introduction of operators such as the right hand side of (2.7) is a
standard technique in the stochastic limit of QFT. We refer to [AcAlFriLu]
for a survey and a detailed discussion, while here we limit ourselves to state
that the choice (2.7) is dictated by the application of first order
perturbation theory to the interaction Hamiltonian considered in Section
(3) below.

Having introduced the collective operators, the
next step of the stochastic approximation is to compute the
2--point function
$$<\Phi, A_\lambda(S,T;f_0,f_1) A^+_\lambda(S',T';f'_0,f'_1)
\Phi>=$$
$$=\lambda^2 \int_{S/\lambda^2} ^{T/\lambda^2}dt\int_{S'/\lambda^2}
^{T'/\lambda^2}dse^{i\omega (t-s)} <\Phi,A(S_tf_0)A(S_tf_1)
A^+(S_sf'_0)A^+(S_sf'_1)\Phi>\eqno(2.8)$$
By the CAR, the scalar product in the right hand side of (2.8) is equal to
$$<S_tf_0,S_sf'_1><S_tf_1,S_sf'_0> -<S_tf_0,S_sf'_0><S_tf_1,S_sf'_1>
\eqno(2.9)$$

By standard arguments [AcLu1] one proves that
the limit, as $\lambda\to 0$, of (2.8) is
$$<\chi_{[S,T]} ,\chi_{[S',T']}>_{L^2({\bf R})}\cdot
\int_{-\infty}^\infty ds e^{i\omega s}
\bigl[<f_0,S_sf'_1><f_1,S_sf'_0> -<f_0,S_sf'_0><f_1,S_sf'_1>
\bigr]\eqno(2.10)$$
Let us introduce on the algebraic tensor product ${\cal K}\otimes{\cal K}$
the pre--scalar product $(\cdot|\cdot)$ defined by:
$$(f_0\otimes f_1 |f'_0\otimes f'_1 ):=$$
$$=\int_{-\infty}^\infty ds e^{i\omega s}
\bigl[<f_0,S_sf'_1><f_1,S_sf'_0> -<f_0,S_sf'_0><f_1,S_sf'_1>
\bigr]\eqno(2.11)$$
and denote by ${\cal K}\otimes_{\rm FB}{\cal K}$ the Hilbert space
obtained by completing ${\cal K}\otimes{\cal K}$ with this scalar product.
That the sesquilinear form (2.11) is positive is not obvious at first
sight, but follows from the fact that (2.10) is the limit of (2.8).
Now let us consider the correlator
$$<\Phi, \prod_{k=1}^nA_\lambda^{\varepsilon
(k)}(S_k,T_k;f_{0,k},f_{1,k})\Phi>\eqno(2.12)$$
where,
$$A^\varepsilon:=\cases{A^+,\ &if $\varepsilon=1$,\cr
A,\ &if $\varepsilon=0$\cr} $$

By the CAR, it is easy to see that
\bigskip
$\underline{\rm {\bf LEMMA}\ The expression (2.1)}$ (2.12) is equal to zero if
the
number of creators is different from the number of annihilators, i.e.
$$\Bigl|\{k;\ \varepsilon (k)=1\}\Bigr|\not=
\Bigl|\{k;\ \varepsilon (k)=0\}\Bigr| \eqno(2.13)$$
or if there exsits a $j=1,\cdots, n$ such that the number of creators on
the left of $j$ is greater than the number of annihilators with the same
property, i.e.
$$\Bigl|\{k\le j;\ \varepsilon (k)=1\}\Bigr|>
\Bigl|\{k\le j;\ \varepsilon (k)=0\}\Bigr| \eqno(2.14)$$
\bigskip

$\underline{\rm {\bf THEOREM}\ (2.2)}$  The limit, as $\lambda\to 0$,
of (2.12) is equal to
$$<\Psi, \prod_{k=1}^na^{\varepsilon
(k)}(\chi_{[S_k,T_k]}\otimes f_{0,k}\otimes_{\rm
FB}f_{1,k})\Psi>\eqno(2.15)$$
where, $a, a^+$ and $\Psi$
are (Boson) annihilation, creation operators on the Boson Fock space
$\Gamma_+(L^2({\bf R})\otimes {\cal K}\otimes_{\rm FB}{\cal K})$ respectively
and $\Psi$ is the vacuum vector.

\bigskip
$\underline{\rm Proof}$ By Lemma (2.1) and the CCR, we need only
to prove the statement in the case $n=2N$ and $\varepsilon$ such that
neither (2.13) nor (2.14) are true. In this case, by the
definition of the collective operators, (2.12) is equal to
$$\lambda^{2N}\int_{S_1/\lambda^2} ^{T_1/\lambda^2}dt_1 \cdots
\int_{S_{2N}/\lambda^2} ^{T_{2N}/\lambda^2}dt_{2N}
e^{i(-1)^{1-\varepsilon(1)}\omega t_1}\cdots
e^{i(-1)^{1-\varepsilon(2N)}\omega t_{2N}}$$
$$<\Phi,A^{\varepsilon(1)}(S_{t_1}f_{\varepsilon(1),1})
A^{\varepsilon(1)}(S_{t_1}f_{1-\varepsilon(1),1}) \cdots
A^{\varepsilon(2N)}(S_{t_{2N}}f_{\varepsilon(2N),2N})
A^{\varepsilon(2N)}(S_{t_{2N}}f_{1-\varepsilon(2N),2N}) \Phi>
\eqno(2.16)$$

By the CAR, the scalar product in (2.16) is equal to
$$\sum_{1\le j_k,p_k,q_k\le n,\ \epsilon,\epsilon'\in \{0,1\}^{2N} \atop
j_k<p_k,j_k<q_k, \varepsilon(j_k)=0,\varepsilon(p_k)=\varepsilon(q_k)=1}
R(\epsilon,\epsilon', \{j_k,p_k,q_k \})$$
$$\prod_{k=1}^N<S_{t_{j_k}}f_{0,j_k}, S_{t_{p_k}}f_{\epsilon(p_k),p_k}>
<S_{t_{j_k}}f_{1,j_k}, S_{t_{q_k}}f_{\epsilon'(q_k),q_k}> =$$
$$=\sum_{1\le j_k<p_k\le n,\ \epsilon,\in \{0,1\}^{2N}}
R(\epsilon,\{j_k,p_k\}) $$
$$\prod_{k=1}^N<S_{t_{j_k}}f_{0,j_k}, S_{t_{p_k}}f_{\epsilon(p_k),p_k}>
<S_{t_{j_k}}f_{1,j_k}, S_{t_{p_k}}f_{1-\epsilon(p_k),p_k}>+$$
$$+\sum_{1\le j_k,p_k\not= q_k\le n,\ \epsilon,\epsilon'\in \{0,1\}^{2N} \atop
j_k<p_k,j_k<q_k, \varepsilon(j_k)=0,\varepsilon(p_k)=\varepsilon(q_k)=1}
R(\epsilon,\epsilon', \{j_k,p_k,q_k \})\cdot$$
$$\cdot\prod_{k=1}^N<S_{t_{j_k}}f_{0,j_k}, S_{t_{p_k}}f_{\epsilon(p_k),p_k}>
<S_{t_{j_k}}f_{1,j_k}, S_{t_{q_k}}f_{\epsilon'(q_k),q_k}>\eqno(2.17)$$
where, $R$ is some power of $-1$.

The second term in the right hand side of (2.17) consists of the terms in
which there exsits $k$, such that the annihilation operators $A(
S_{t_{j_k}}f_{0,j_k} )$, $A(S_{t_{j_k}}f_{1,j_k})$ are used to produce a
scalar product with the creation operators $A^+(S_{t_{p_k}}f_{\epsilon(p_k),
p_k})$, $A^+(S_{t_{q_k}}f_{\epsilon(q_k),q_k})$ respectively with
$p_k\not= q_k$. Such terms will be called
{\bf cross--terms}.

Since the absolute value of the second term in the right hand side of
(2.17) is less than or equal to
$$\sum_{1\le j_k,p_k\not= q_k\le n,\ \epsilon,\epsilon'\in \{0,1\}^{2N} \atop
j_k<p_k,j_k<q_k, \varepsilon(j_k)=0,\varepsilon(p_k)=\varepsilon(q_k)=1}
\prod_{k=1}^N|<S_{t_{j_k}}f_{0,j_k}, S_{t_{p_k}}f_{\epsilon(p_k),p_k}>
<S_{t_{j_k}}f_{1,j_k}, S_{t_{q_k}}f_{\epsilon'(q_k),q_k}>|\eqno(2.18)$$
Lemma (3.2) in [Lu] guarantees that this term contributes to our limit
trivially, i.e. the limit of the scalar product in (3.10) can be
replaced by the first term of the right hand side of (2.17).

Now, we compute the limit of the
first term of the right hand side of (2.17). For
each $1\le j_k<p_k\le n$, consider the quantity
$$<\Phi, \cdots A(S_{t_{j_k}}f_{0,j_k} )A(S_{t_{j_k}}f_{1,j_k})
\cdots A(S_{t_{p_k}}f_{1,p_k} )A(S_{t_{p_k}}f_{0,p_k})
\cdots\Phi>\eqno(2.19)$$
in which the annihilation operators $A(S_{t_{j_k}}f_{0,j_k} )$,
$A(S_{t_{j_k}}f_{1,j_k})$ are used to produce
scalar products with the creation operators $A^+(S_{t_{p_k}}f_{1,p_k})$,
$A^+(S_{t_{p_k}}f_{0,p_k})$. In order to do this, the annihilation operators
$A(S_{t_{j_k}}f_{0,j_k} )$, $A(S_{t_{j_k}}f_{1,j_k})$ must be on
the left of the creation operator
$A^+(S_{t_{p_k}}f_{1,p_k})$. This procedure, by the CAR, gives a factor
$$(-1)^{2(p_k-j_k-1)}\cdot (-1)^{2(p_k-j_k-1)}=1$$
(the appearance of 2 is due to the fact that there are
two operators corresponding to each time moment).
By doing this for any pair $j_k<p_k$, it follows from the CAR that the
first term of the right hand side of (2.10) is equal to
$$\sum_{1\le j_k<p_k\le 2N}\prod_{k=1}^N \lambda^2
\int_{S_{j_k}/\lambda^2} ^{T_{j_k}/\lambda^2}dt_{j_k}
\int_{S_{p_k}/\lambda^2} ^{T_{p_k}/\lambda^2}dt_{p_k}$$
$$<\Phi, A(S_{t_{j_k}}f_{0,j_k} )A(S_{t_{j_k}}f_{1,j_k})
A(S_{t_{p_k}}f_{1,p_k} )A(S_{t_{p_k}}f_{0,p_k})\Phi>\eqno(2.20)$$
By sending $\lambda$ to zero, we complete the proof using (2.8) and
(2.10).\bigskip\bigskip

\beginsection{\S3.) Stochastic Bosonization for interacting systems}

\bigskip
We shall now apply the result of the previous Section to
the following interaction Hamiltonian:
$$H_I:=i\lambda (D\otimes A^+_Q(g_0)A^+_Q(g_1) -D^+\otimes A_Q(g_1)A_Q(g_0))
\eqno(3.1)$$
where $\lambda>0$ is a real number, $D$ is a bounded (but this is not
really necessary) operator on a space $H_S$, called {\it the system
space\/} and $A_Q$, $A^+_Q$ are creation and annihilation operators
which we suppose realized on a representation space of the CAR--algebra
over a 1--particle space ${\cal H}_1$. The initial state is given by
$$<\Phi_Q,\cdot\ \Phi_Q>\eqno(3.2)$$
with $\Phi$ cyclic for the representation and Gaussian with 2--point
function
$$<\Phi_Q,A^+_Q(f)A_Q(g)\Phi_Q>=<f,{1-Q\over2}g>\eqno(3.3)$$
$\Gamma_-({\cal H}_1)$
where $Q$ 1 is an operator on ${\cal H}_1$, for example
$Q={1-ze^{-\beta H_1}\over1+ze^{-\beta H_1}}$.

It is well known that up to a unitary isomorphism
$$A_Q(f)=A(Q_+f)\otimes 1+\theta \otimes A^+(\iota Q_-f)\eqno(3.4)$$
where is the Fermi Fock space and
the right hand side of (3.4) is an operator on the Hilbert space
$$\Gamma_-({\cal H}) \otimes \Gamma_-({\cal H}_\iota)$$
where ${\cal H}_\iota$ is the conjugate space of ${\cal H}$ (the space
of {\it bra\/} vectors) and
$$Q_+=\sqrt{1+Q\over2} ,\qquad Q_-=\sqrt{1-Q\over2} ,\qquad
\theta:=\bigoplus_{n=0}^\infty(-1)^n\eqno(3.5)$$
Thus, up to isomorphism,
$$A_Q(g_1)A_Q(g_0) =A(Q_+g_1)A(Q_+g_0)\otimes1+ A(Q_+g_1)\theta\otimes
A(\iota Q_-g_0)+ $$
$$+\theta A(Q_+g_0)\otimes A(\iota Q_-g_1) +
1\otimes A(\iota Q_-g_0)A(\iota Q_-g_1)\eqno(3.6)$$
By considering $2n$--point functions, one can prove that, second the
third and fourth
terms give trivial contribution to our WCL. So the problem is reduced
to Fock case.


Thanks to the discussion in \S1, we can limit ourselves to the Fock case.
In this case, we have:

i) the Fermi--Fock space $\Gamma_-({\cal H}_1)$,

ii) the interaction Hamiltonian
$$H_I:=i\lambda (D\otimes A^+(g_0)A^+(g_1) -D^+\otimes A(g_1)A(g_0)
)\eqno(3.7)$$

iii) the vacuum vector $\Phi:=1\oplus 0\oplus 0\oplus\cdots$,

iv) a unitary group $\{S_t:=e^{iH_1t}\}_{t\ge0}$ on the Hilbert space
${\cal H}_1$ with a self--adjoint operator $H_1$,

v) a subspace ${\cal K}$ of ${\cal H}$  with the property
$$\int_{-\infty}^\infty dt|<f,S_tg>|<\infty ,\qquad\forall f,g\in
{\cal K} \eqno(3.8)$$

vi) the evolved interaction Hamiltonian
$$H_I(t):=e^{i(H_0\otimes 1+1\otimes d\Gamma(H_1))t} H_I
e^{-i(H_0\otimes 1+1\otimes d\Gamma(H_1))t} \eqno(3.9)$$

vii) the evolution operators $\{U^{(\lambda)}_t\}_{t\ge0}$ defined as the
solution of Sch\"odinger equation:
$${d\over dt}U^{(\lambda)}_t=-i\lambda H_I(t)U^{(\lambda)}_t \
,\qquad\qquad U^{(\lambda)}_0 =1\eqno(3.10)$$
where, $\lambda$ is the coupling constant.

$U^{(\lambda)}_t$ has the form
$$U^{(\lambda)}_t =1+\sum_{n=1}^\infty \int_0^tdt_1\int_0^{t_1}dt_2
\cdots \int_0^{t_{n-1}}dt_n (-i\lambda)^n H(t_1) \cdots H(t_n)
\eqno(3.11)$$
which is weakly convergent on appropriate domain.

Our basic assumption is that the system Hamiltonian has discrete
spectrum or, at least, that it has a complete orthonormal basis $(e_n)$
of eigenvectors. Thus we have that
$$D(t):=e^{iH_0t} De^{-iH_0t} =\sum_{n,m=0}^\infty e^{ix_nt} |e_n><e_n|
D|e_m><e_m| e^{-ix_mt}  =$$
$$=\sum_{n,m=0}^\infty e^{i(x_n-x_m)t} |e_n><e_n| D|e_m><e_m| \eqno(3.12)$$
Denote by $\{\omega_j\}_{j=0}^\infty$ the set $\{x_n-x_m;\
n,m=0,1,2,\cdots\}$, then $D(t)$ has the form
$$\sum_{j=0}^\infty e^{i\omega_j t}D_j\eqno(3.13)$$
where
$$D_j:=\sum_{n,m\atop x_n-x_m =\omega_j}|e_n><e_n|D|e_m><e_m|
\eqno(3.14)$$
Notice that, with this notations we have
the property: $\omega_j\not=\omega_r$ if $j\not= r$.

The limit of $U_{t/\lambda^2}^{(\lambda)}$, as $\lambda\to 0$ will be
investigated in the present note.

We shall consider first of all the case of non--rotating wave
approximation, i.e.
$$D(t)=e^{i\omega t} D\eqno(3.15)$$
then by combining the results and techniques of the present paper with
techniques now standard in the stochastic limit of quantum field theory,
we shall state the result in the general case.
\bigskip\bigskip

\noindent{\bf \S4.)\ The limit process}

\bigskip
In this Section we investigate the limit, as $\lambda \to 0$, of
matrix elements (with respect to the
collective number vectors) of the wave operator
$U^{(\lambda)}_t$ i.e., expanding $U^{(\lambda)}_t$ in series:
$$<\prod_{k=1}^NA_\lambda^+(S_k,T_k;f_{0,k},f_{1,k}) \Phi,
\sum_{n=0}^\infty \lambda^n\int_0^{t/\lambda^2}dt_1\int_0^{t_1}dt_2
\cdots \int_0^{t_{n-1}}dt_n$$
$$\prod_{k=1}^n A^{\varepsilon(k)}(S_{t_{k}}g_{1-\varepsilon(k)})
A^{\varepsilon(k)}(S_{t_{k}}g_{\varepsilon(k)})e^{i(-1)^{1-\varepsilon(k)}
\omega t_k}
\prod_{k=1}^{N'}A_\lambda^+(S'_k,T'_k;f'_{0,k},f'_{1,k})\Phi>\eqno(4.1)$$

First of all, since the difference between the CCR and the CAR in a scalar
product such as (4.1) consists only of some power of $-1$,
the uniform estimate of Lemma (3.2) of [Lu] can be applied directly and
we have that

\bigskip
$\underline{\rm {\bf THEOREM}\ (4.1)}$ The limit  in the sum over $n$ of
the terms (4.1) can be
performed term by term.\bigskip

By using the notations, (2.2) and (2.7), we see the (4.1) can
be rewritten as
$$\sum_{n=0}^\infty\lambda^{n+N+N'}
\int_{S_{1}/\lambda^2} ^{T_{1}/\lambda^2}du_{1}
\cdots \int_{S_{N}/\lambda^2} ^{T_{N}/\lambda^2}du_{N}$$
$$\int_{S'_{1}/\lambda^2} ^{T'_{1}/\lambda^2}dv_{1}
\cdots \int_{S'_{N'}/\lambda^2} ^{T'_{N'}/\lambda^2}dv_{N'}
\int_0^{t/\lambda^2}dt_1\int_0^{t_1}dt_2
\cdots \int_0^{t_{n-1}}dt_n$$
$$e^{i\omega(\sum_{k=1}^{N'}v_k+\sum_{k=1}^n(-1)^{1-\varepsilon(k)}
t_k-\sum_{k=1}^{N}u_k)}$$
$$< \Phi,
\bigl[\prod_{k=1}^N{\cal A}^+(S_{u_k}f_{0,k},S_{u_k}f_{1,k}) \bigr]^*
\prod_{k=1}^n {\cal A}^{\varepsilon(k)}(S_{t_{k}}g_{\varepsilon(k)}),
S_{t_{k}}g_{1-\varepsilon(k)})
\prod_{k=1}^{N'}{\cal A}^+(S_{v_k}f'_{0,k},S_{v_k}f'_{1,k})\Phi>
\eqno(4.7)$$

Now we shall, by applying the quasi--CCR, calculate the scalar product
in (4.7). Suppose that the annihilation operator on the extreme right in the
product of operators in (4.7) corresponds
to the time--moment $s$ and look at the product
$${\cal A}(S_{s}\bar g_{0},S_{s}\bar g_{1}) {\cal A}^+(S_{\tau}g'_{0},
S_{\tau}g'_{1}) \eqno(4.8)$$
where, in the notations of (4.7):

i) if $s=u_1$, then $\tau$ is defined as $t_1$ and
$\bar g_\varepsilon:=f_{\varepsilon,1} $,
$g'_\varepsilon:=g_{\varepsilon}$;

ii) if $s=t_p$ for some $p\le n-1$, then $\tau$ is defined as
$t_{p+1}$ and, in the notations of i), $\bar g_\varepsilon:=g_{\varepsilon}$,
$g'_\varepsilon:=g_{\varepsilon}$;

iii) if $s=t_n$, then $\tau$ is defined as $v_1$ and
$\bar g_\varepsilon:=g_{\varepsilon}$,
$g'_\varepsilon:=f_{\varepsilon,1}$.

By the quasi--CCR, (4.8) is equal to
$$ <(S_{s}\bar g_{0},S_{s}\bar g_{1}),(S_{\tau}g'_{0},S_{\tau}g'_{1})>
+{\cal A}^+(S_{s}\bar g_{0},S_{s}\bar g_{1})
{\cal A}(S_{\tau}g'_{0},S_{\tau}g'_{1}) +$$
$$+R(S_{s}\bar g_{0},S_{s}\bar g_{1};S_{\tau}g'_{0},S_{\tau}g'_{1})
\eqno(4.9)$$
Hence the scalar product in (4.7) is equal to:
$$<\Phi, {\rm a\ product\ of \ annihilation\ and\ creation\
operators }\cdot$$
$$\cdot\Bigl(<(S_{s}\bar g_{0},S_{s}\bar
g_{1}),(S_{\tau}g'_{0},S_{\tau}g'_{1})>
+{\cal A}^+(S_{s}\bar g_{0},S_{s}\bar g_{1})
{\cal A}(S_{\tau}g'_{0},S_{\tau}g'_{1})+ $$
$$+R(S_{s}\bar g_{0},S_{s}\bar g_{1};S_{\tau}g'_{0},S_{\tau}g'_{1})
\Bigr)\cdot{\rm a\ product\ of \ creation\ operators }\cdot\Phi>\eqno(4.10)$$

Let us study the third term in (4.10), i.e.:
$$<\Phi, {\rm a\ product\ of \ annihilation\ and\ creation\
operators }\cdot$$
$$\cdot R(S_{s}\bar g_{0},S_{s}\bar g_{1};S_{\tau}g'_{0},S_{\tau}g'_{1})\cdot\
{\rm a product of creation operators}\ \Phi>\eqno(4.11)$$

Notice that $R(S_{s}\bar g_{0},S_{s}\bar
g_{1};S_{\tau}g'_{0},S_{\tau}g'_{1})$ is an operator equal to a
sum of three quantities of the form $<S_sf,S_\tau f'>A^+(S_\tau g)
A(S_sg')$. Therefore (4.11) is equal to a sum of three
quantities like
$$<\Phi, {\rm a\ product\ of \ annihilation\ and\ creation\ operators }
\ A^+(S_\tau g)A(S_sg)$$
$${\rm a\ product\ of \ creation\ operators }\ \Phi><S_sf,S_\tau
f'>\eqno(4.12)$$
It is certainly true that the annihilation (resp. creation) operator
$A(S_sg')$ (resp. $A^+(S_\tau g)$) must be used to produce a scalar product
with a creation operator $A^+(S_u h')$ (resp. $A^+(S_v h)$)
with $u\not= s$ (resp. $v\not=\tau $). In other words, (4.12) is
equal to a sum of cross--terms and therefore contributes to our limit
only zero. That is, up to an $o(1)$, (4.10) is equal to
$$<\Phi, {\rm a\ product\ of \ annihilation\ and\ creation\
operators}\ \cdot$$
$$\Bigl(<(S_{s}\bar g_{0},S_{s}\bar g_{1}),(S_{\tau}g'_{0},S_{\tau}g'_{1})>
+{\cal A}^+(S_{s}\bar g_{0},S_{s}\bar g_{1})
{\cal A}(S_{\tau}g'_{0},S_{\tau}g'_{1}) \Bigr)\cdot$$
$${\rm a\ product\ of \ creation\ operators}\ \Phi>\eqno(4.13)$$
Repeating the above discussion, we find that although ${\cal A}^{\pm}$ are
not Boson objects, as $\lambda\to 0$, the remainder $R$ can be
effectively forgotten and they satisfy the CCR. Therefore, the
following result can be obtained by combining the above discussions and
the techniques in [Lu]:

\bigskip
$\underline{\rm {\bf THEOREM}\ (4.2)}$ For any $\xi,\eta \in {\cal
H}_0$, the limit, as $\lambda\to 0$, of
$$<\xi\otimes\prod_{k=1}^NA_\lambda^+(S_k,T_k;f_{0,k},f_{1,k}) \Phi,
U^{(\lambda)}_{t/\lambda^2}\eta\otimes
\prod_{k=1}^{N'}A_\lambda^+(S'_k,T'_k;f'_{0,k},f'_{1,k})  \Phi>
\eqno(4.13)$$
exsits and is equal to
$$<\xi\otimes\prod_{k=1}^Na^+(\chi_{[S_k,T_k]}\otimes f_{0,k}\otimes
_{\rm FB}f_{1,k}) \Psi,U(t)
\prod_{k=1}^{N'}a^+(\chi_{[S'_k,T'_k]}\otimes f'_{0,k}\otimes
_{\rm FB}f'_{1,k}) \Psi>\eqno(4.14)$$
where, $U(t)$ is the solution of the Boson stochastic differential
equation
$$U(t)=1+\int_0^t\Bigl(D\otimes da^+_s(g_0\otimes_{\rm FB}g_1) -
D^+\otimes da_s(g_0\otimes_{\rm FB}g_1) -$$
$$-D^+D\otimes 1
(g_0\otimes_{\rm FB}g_1|g_0\otimes_{\rm FB}g_1)_-
\Bigr)U(s)\eqno(4.15)$$
and where the half scalar product is defined as
$$(f\otimes_{\rm FB}g|f'\otimes_{\rm FB}g')_- :=
\int_{-\infty}^0dt e^{i\omega t}\bigl(<f, S_tf'><g,S_tg'>-
<f, S_tg'><g,S_tf'> \bigr)\eqno(4.16)$$
Moreover, $\{U(t)\}_{t\ge0}$ is a unitary process.
\vfill\eject

\noindent{\bf \S5.)\ Discrete spectrum case}

\bigskip
Now we discuss the problem in the more general case in which one
replaces the assumption (3.9) by (3.6).
In this case, the collective creation operators are defined as:
$$A^+_\lambda(S,T;\{f^{(j)}_0,f^{(j)}_1\}):=\sum_{j\in {\bf Z}}
\lambda\int_{S/\lambda^2} ^{T/\lambda^2}
e^{i\omega_j t} A^+(S_tf^{(j)}_0)A^+(S_tf^{(j)}_1)dt\eqno(5.1)$$
for $S\le T$ and $f^{(j)}_0,f^{(j)}_1\in {\cal K}$ with the property
that there are only a finite number of $j$ such that $f^{(j)}_0\not=0,
f^{(j)}_1\not=0$. The analogue of Theorem (3.2) will then be

\bigskip
$\underline{\rm {\bf THEOREM}\ (5.1)}$  The limit, as $\lambda\to 0$,
of
$$<\Phi, \prod_{k=1}^nA_\lambda^{\varepsilon
(k)}(S_k,T_k;f^{(j)}_{0,k},f^{(j)}_{1,k})\Phi>\eqno(5.2)$$
exsits and equal to
$$<\Psi, \prod_{k=1}^na^{\varepsilon
(k)}(\chi_{[S_k,T_k]}\otimes \bigoplus_{j\in{\bf Z}}
f^{(j)}_{0,k}\otimes_{\rm FB}f^{(j)}_{1,k})\Psi>\eqno(5.3)$$
where, $a, a^+$ and $\Psi$
are (the Boson) annihilation, creation operators on the Boson--Fock
space $\Gamma_+(L^2({\bf R})\otimes \bigoplus_{j\in {\bf Z}}
{\cal K}\otimes_{\rm FB}^{(j)}{\cal K})$ respectively
and $\Psi$ is the vacuum vector and for each $j\in {\bf Z}$
$$(f_0\otimes_{\rm FB}^{(j)} f_1 |f'_0\otimes_{\rm FB}^{(j)} f'_1 ):=$$
$$=\int_{-\infty}^\infty ds e^{i\omega_j s}
\bigl[<f_0,S_sf'_1><f_1,S_sf'_0> -<f_0,S_sf'_0><f_1,S_sf'_1>
\bigr]\eqno(5.4)$$
\bigskip
The final result is

\bigskip
$\underline{\rm {\bf THEOREM}\ (5.2)}$ For any $\xi,\eta \in {\cal
H}_0$, the limit, as $\lambda\to 0$, of
$$<\xi\otimes\prod_{k=1}^NA_\lambda^+(S_k,T_k;\{f^{(j)}_{0,k},f^{(j)}_{1,k}\})
\Phi,U^{(\lambda)}_{t/\lambda^2}\eta\otimes
\prod_{k=1}^{N'}A_\lambda^+(S'_k,T'_k;\{h^{(j)}_{0,k},h^{(j)}_{1,k})  \Phi>
\eqno(5.5)$$
exsits and is equal to
$$<\xi\otimes\prod_{k=1}^Na^+(\chi_{[S_k,T_k]}\otimes
\bigoplus_{j\in{\bf Z}}f^{(j)}_{0,k}\otimes^{(j)} _{\rm FB}f^{(j)}_{1,k})
\Psi,$$
$$U(t)\prod_{k=1}^{N'}a^+(\chi_{[S'_k,T'_k]}\otimes
\bigoplus_{j\in{\bf Z}}h^{(j)}_{0,k}\otimes
^{(j)}_{\rm FB}h^{(j)}_{1,k}) \Psi>\eqno(5.6)$$
where, $U(t)$ is the solution of the Boson stochastic differential
equation
$$U(t)=1+\int_0^t\sum_{j\in{\bf Z}}\Bigl(D_j\otimes da^+_s(
0\bigoplus g_0\otimes^{(j)}_{\rm FB}g_1\bigoplus 0) -D_j^+\otimes
da_s(0\bigoplus g_0\otimes^{(j)}_{\rm FB}g_1\bigoplus 0) -$$
$$-D_j^+D_j\otimes 1
(g_0\otimes^{(j)}_{\rm FB}g_1|g_0\otimes^{(j)}_{\rm FB}g_1)_-
\Bigr)U(s)\eqno(5.7)$$
and the half scalar product is defined as
$$(f\otimes^{(j)}_{\rm FB}g|f'\otimes^{(j)}_{\rm FB}g')_- :=
\int_{-\infty}^0dt e^{i\omega_j t}\bigl(<f, S_tf'><g,S_tg'>-
<f, S_tg'><g,S_tf'> \bigr)\eqno(5.8)$$
Moreover, $\{U(t)\}_{t\ge0}$ is a unitary process.
\bigskip
$\underline{\rm {\bf Proof}}$ The theorem can proved by
combining the results in the present note with those in [Lu].
\bigskip\bigskip

\noindent{\bf Aknowledgements}\bigskip

L.A. acknowledges partial support from the Human Capital and Mobility
programme, contract number: erbchrxct930094.\bigskip\bigskip
\noindent{\bf References}\bigskip

\item{[AcLuVo93]}
L. Accardi, Y.G.Lu, I. Volovich:
The Stochastic Sector of Quantum Field Theory.
Volterra Preprint N.138, 1993;
Matematicheskie Zametki (1994)
\item{[AcLuVo94b]} Accardi L., Lu Y.G., Volovich I.:
Non--Commutative (Quantum) Probability, Master Fields and Stochastic
Bosonization
Preprint Volterra (1994) ,hep-th/9412241.
\item{[AcFrLu87]} Accardi L., Frigerio A., Lu Y.G.:
On the weak coupling limit problem
in : Quantum Probability and Applications IV
Springer LNM N. 1396(1987)20-58
\item{[AcFrLu90]} Accardi L., Frigerio A., Lu Y.G.:
The weak coupling limit as a quantum functional central limit,
Comm. Math. Phys. 131 (1990) 537-570
\item{[AcLu]} Accardi L.,  Lu Y.G.:
Quantum Electro Dynamics: the master and the Langevin equation.
invited talk at the Symposium
in: {\it Mathematical Approach to Fluctuations: Astronomy, Biology and
Quantum Dynamics }; Vol. I, Proceedings of the
Kyoto International Institute for Advanced Studies, May 18-22, 1992, ed.
T. Hida, World Scientific (1994) 3--22
Volterra preprint N.133 (1993)
\item{ [AcAlFriLu]} Accardi L., Alicki R., Frigerio A., Lu Y.G.:
An invitation to the weak coupling and the low density limit.
in: Quantum Probability and Applications VI (1989)
\item{[Lu]} Lu Y.G.
The weak coupling limit for quadratic interactions
J. Math. Phys. {\bf 33}, 8, (1992).
\item{[AcLuVo94a]} L. Accardi, Yun Gang Lu, Igor Volovich:
On the stochastic limit of Quantum Chromodynamics,
to appear in: Quantum Probability and Related Topics, World Scientific,
OP--PQ IX (1994)
\item{[Lut]} A. Luther, {\it Phys. Rev.\/}, {\bf B19} (1979) 320.
\item{[Hal]} F.D.M. Haldane, {\it Helv. Phys. Acta\/}, {\bf 65} (1992) 152.
\item{[SV]} V.N. Sushko and I.V. Volovich, ``Constructive quantum field
theory. Thirring model local field'', {\it Theoret. Mathem. Phys.\/},
{\bf 9} (1972) 136--151.
\item{[Wh]} A. Wightman, {\it Introduction to some aspects of the
relativistic dynamics of quantized fields\/}, in: ``High Energy
Electromagnetic Interactions and Field Theory'', M. Levy ed., Now York:
Gordon and Breach.\bye